\documentclass[amsmath,amssymb,amsfonts,aps,pra,twocolumn,showpacs]{revtex4-1}
\usepackage{times}
\usepackage[pdftex]{graphicx}
\usepackage[utf8]{inputenc}
\usepackage[urlcolor=blue, hyperindex, colorlinks, bookmarks=true, linkcolor=blue, citecolor=blue]{hyperref}

\begin{document}

\title{Detection of weak forces based on noise-activated switching in bistable optomechanical systems}
\author{Samuel~Aldana}
\author{Christoph~Bruder}
\author{Andreas~Nunnenkamp}
\affiliation{Department of Physics, University of Basel, Klingelbergstrasse 82, CH-4056 Basel, Switzerland}

\date{\today}

\begin{abstract}
We propose to use cavity optomechanical systems in the regime of optical
bistability for the detection of weak harmonic forces. Due to the
optomechanical coupling an external force on the mechanical oscillator
modulates the resonance frequency of the cavity and consequently the switching
rates between the two bistable branches. A large difference in the cavity
output fields then leads to a strongly amplified homodyne signal. We determine
the switching rates as a function of the cavity detuning from extensive
numerical simulations of the stochastic master equation as appropriate for
continuous homodyne detection. We develop a two-state rate equation model that
quantitatively describes the slow switching dynamics. This model is solved
analytically in the presence of a weak harmonic force to obtain approximate
expressions for the power gain and signal-to-noise ratio that we then compare
to force detection with an optomechanical system in the linear regime.
\end{abstract}


\maketitle

\section{Introduction}

The field of cavity optomechanics is historically closely related to the problem
of force sensing in the context of gravitational wave detection
\cite{Braginsky1968,Braginsky1975,Caves1980a,Caves1980b}, and the fundamental limit
of force sensitivity can be traced back to the quantum-mechanical nature of the
detector, the so-called standard quantum limit \cite{Braginsky1992}.

Most optomechanical devices to date operate in the regime where the radiation
pressure is sufficiently weak on the single-photon level so the coupling
between phonons and photons can be linearized. Examples for exciting
progress in this area include the observation of ground-state cooling
\cite{Teufel2011cooling, Riviere2011, Chan2011}, ponderomotive squeezing
\cite{Brooks2012,Safavi-Naeini2013,Purdy2013squeezing}, radiation-pressure
shot-noise \cite{Murch2008,Purdy2013rpsn}, and mechanical zero-point motion via
sideband thermometry \cite{Safavi-Naeini2012,Brahms2012,Lee2014,Purdy2014}
as well as the demonstration of displacement detection close to the standard
quantum limit \cite{Anetsberger2009,Schliesser2009,Teufel2009,Schreppler2014}.

Advances in fabricating optomechanical devices promise increasingly large
coupling strengths \cite{Chan2012} making nonlinear quantum effects
\cite{Nunnenkamp2011, Rabl2011} a possible reality in the near future. It is
thus of great interest to study how the intrinsically nonlinear radiation
pressure can be exploited in novel devices.

In this paper we propose sensitive force detection exploiting optical
bistability in an optomechanical system \cite{Dorsel1983, Meystre1985,
Aldana2013}. The optomechanical system we consider consists of a laser-driven
optical cavity whose resonance frequency is modulated by the displacement of a
mechanical oscillator \cite{Kippenberg2008, Marquardt2009, Aspelmeyer2013}.
Under certain conditions the system exhibits an optical bistability, i.e.~it
has two classically stable states with potentially largely different cavity
fields. Shot-noise fluctuations in the coherent drive of the cavity will cause
transitions between the two branches whose switching rates can depend strongly
on cavity detuning. A weak periodic forcing of the mechanical resonator will
modulate the cavity detuning and thus the switching rates allowing the
detection of weak forces in the cavity spectrum.

Note that exploiting periodic modulation of switching rates in bistable systems
to detect small coherent signals has also been discussed in the context of
stochastic resonance \cite{Gammaitoni1998, Wellens2004, Venstra2013} and
Josephson bifurcation amplifiers \cite{Manucharyan2007, Mallet2009, Vijay2009}.

In the following we calculate numerically the switching dynamics in the
single-photon strong-coupling regime and zero temperature limit using a
stochastic quantum master equation. We obtain the switching rates and their
dependence on the detuning from the residence time distribution. We then
develop a two-state rate equation model allowing us to write the output
spectral density of the amplitude quadrature as the sum of a low-frequency
noise background and a signal peak caused by the weak harmonic force.
The homodyne signal amplitude depends linearly on the force amplitude
and on the difference between the cavity output fields.  Bistable
optomechanical systems can thus be used as linear amplifiers whose
bandwidth is the switching rate and which have a potentially large
gain for low-frequency signals.

The remainder of the present paper is organized as follows. In
Sec.~\ref{sec:model} we introduce the model for an optomechanical system (OMS)
with an additional external force driving the mechanical oscillator and present
the stochastic master equation describing the system state conditioned on
continuous homodyne detection. In Sec.~\ref{sec:switching} we investigate
numerically noise-induced switching in a bistable OMS. We obtain time traces of
the homodyne photocurrent, the residence time distributions, and the
switching rates as a function of cavity detuning. In
Sec.~\ref{sec:twostatemodel} we describe the slow switching dynamics and the
influence of a harmonic force within a two-state rate equation model with
periodically modulated switching rates. In Sec.~\ref{sec:forcedetection} we
find expressions for the noise spectral density and the signal amplitude of the
homodyne photocurrent, based on the two-state rate equation model, and compare
them to quantum trajectory results. Finally, we compare the power gain and
signal-to-noise ratio of force detection with a bistable OMS to those
achievable with an OMS in the linear regime.

\section{Model}
\label{sec:model}

We consider an optomechanical system (OMS) in which the position of a
mechanical oscillator modulates the resonance frequency of an optical cavity.
The system consists of a mechanical mode with resonance frequency $\omega_m$
and an optical mode with frequency $\omega_c$ which are coupled by the
radiation-pressure interaction. The optical mode is driven by a laser with
strength $\epsilon$ and frequency $\omega_d$. In a frame rotating at the drive
frequency $\omega_d$ the Hamiltonian reads $(\hbar=1)$
\begin{equation}
\hat{H} = -\Delta_0 \hat{a}^\dagger \hat{a}  -i \epsilon \left( \hat{a} -
\hat{a}^\dagger \right) + \omega_m \hat{b}^\dagger \hat{b} - g_0
\hat{a}^\dagger\hat{a} \left( \hat{b} + \hat{b}^\dagger \right)\:,
\label{eq:hamiltonian}
\end{equation}
where $\hat{a}$ and $\hat{b}$ are bosonic annihilation operators for the optical
and mechanical mode, $\Delta_0 =\omega_d -\omega_c$ is the detuning between
driving and cavity frequency, and $g_0$ is the optomechanical coupling. We also
add an external periodic force on the mechanical resonator with amplitude $g_1$
and frequency $\Omega$
\begin{equation}
\hat{H}_F = -g_1 \sin(\Omega t) \left(\hat{b}+ \hat{b}^\dagger \right)\:.
\label{eq:force}
\end{equation}
A complete description of the system additionally requires the optical damping
rate $\kappa$, the mechanical energy dissipation rate $\gamma_m$, and the mean
phonon number in thermal equilibrium $n_{\text{th}} = 0$ corresponding to a
zero-temperature reservoir.

The dissipative dynamics of the OMS undergoing continuous homodyne measurement
of the cavity output can be described with the It\^o stochastic master
equation (SME) \cite{Wiseman1993,Wiseman2009}
\begin{align}
&d\hat{\rho}_c = \mathcal{L}[\hat{\rho}_c] dt + \mathcal{H}[\hat{\rho}_c] dW\:,
\label{eq:SME}
\\
&\mathcal{L}[\hat{\rho}_c] = -i \left[\hat{H} + \hat{H}_F, \hat{\rho}_c \right]
+ \kappa \mathcal{D}_{\hat{a}} [\hat{\rho}_c]
\nonumber
\\
& \qquad\qquad
+ \left( n_\text{th} + 1 \right) \gamma_m \mathcal{D}_{\hat{b}} [\hat{\rho}_c]
+ n_\text{th} \gamma_m \mathcal{D}_{\hat{b}^\dagger} [\hat{\rho}_c]\:,\\
&\mathcal{H}[\hat{\rho}_c] = \sqrt\kappa \left( \hat{a} \hat{\rho}_c +
\hat{\rho}_c \hat{a}^\dagger - \left\langle  \hat{a}+\hat{a}^\dagger
\right\rangle_c \hat{\rho}_c \right)\:,
\end{align}
where $d\rho_c = \hat{\rho}_c(t+dt)-\hat{\rho}_c(t)$, $\langle
\hat{a}+\hat{a}^\dagger \rangle_c =
\text{Tr}[(\hat{a}+\hat{a}^\dagger)\hat{\rho}_c]$, and $dW$ is a Wiener
increment with $\text{E}[dW]=0$ and $\text{E}[dW^2] = dt$. $\text{E}[-]$ is the
ensemble average and the Lindblad terms have the usual form,
$\mathcal{D}_{\hat{o}} [\hat{\rho}] = \hat{o} \hat{\rho} \hat{o}^\dagger - (
\hat{o}^\dagger\hat{o} \hat{\rho} +  \hat{\rho} \hat{o}^\dagger\hat{o})/2$. The
first term in Eq.~\eqref{eq:SME} is the Liouvillian describing the coherent
evolution due to the Hamiltonian and the decoherence originating from the
coupling to the environment. The second term called innovation describes the
effect of a measurement of the amplitude quadrature, $\hat{X} =
\hat{a}+\hat{a}^\dagger$, with homodyne detection of the cavity output field.
The innovation term conditions the evolution of the quantum state
$\hat{\rho}_c(t)$ on the homodyne photocurrent
\begin{equation}
I_c(t) = \sqrt{\kappa} \big\langle \hat{X}(t) \big\rangle_c + \frac{dW}{dt}\:,
\end{equation}
which is the sum of a conditioned expectation value of $\hat{X}$ and a
fluctuating term originating from the shot noise of the local oscillator (here
we have assumed unit detection efficiency).

We will refer to the result for a particular noise realization of
$\hat{\rho}_c(t)$ and $I_c(t)$ as a quantum trajectory. Taking the ensemble
average of Eq.~\eqref{eq:SME} we recover the unconditional quantum state
$\hat{\rho}(t) = \text{E}[\hat{\rho}_c(t)]$ which is a solution to the quantum
master equation
\begin{equation}
\dot{\hat{\rho}} = \mathcal{L}[\hat{\rho}]\:.
\label{eq:QME}
\end{equation}

In the following we calculate the evolution of the quantum state
$\hat{\rho}_c(t)$ by numerically integrating Eq.~\eqref{eq:SME}
\cite{Kloeden1992} and use the time traces of the homodyne photocurrent
$I_c(t)$ to investigate the switching dynamics in the regime of optical
bistability.

To quantify the influence of the external mechanical force on the cavity output
we use the time-averaged spectral density
\begin{equation}
S_{II}^\text{out}(\omega) = \lim_{t\to\infty} \int d\tau \, e^{i\omega \tau}
\text{E}\left[ I_c(t+\tau) I_c(t) \right]\:.
\end{equation}
For finite, but sufficiently long sampling times $T$ the spectral density can be
obtained using the Wiener-Khintschin theorem from a quantum trajectory as
$S_{II}^\text{out}(\omega) = | I_T(\omega)|^2$ where
\begin{equation}
I_T(\omega) = \frac{1}{\sqrt{T}} \int_0^T dt \, e^{i\omega t} I_c(t)
\end{equation}
is the windowed Fourier transform of the homodyne photocurrent $I_c(t)$. In
this way we replace the ensemble average by a time average.  In the following
we will numerically simulate a single, sufficiently long quantum trajectory
instead of calculating averages over an ensemble of quantum trajectories.

\section{Noise-activated switching in bistable OMS}
\label{sec:switching}

\begin{figure}
\includegraphics[width=\linewidth]{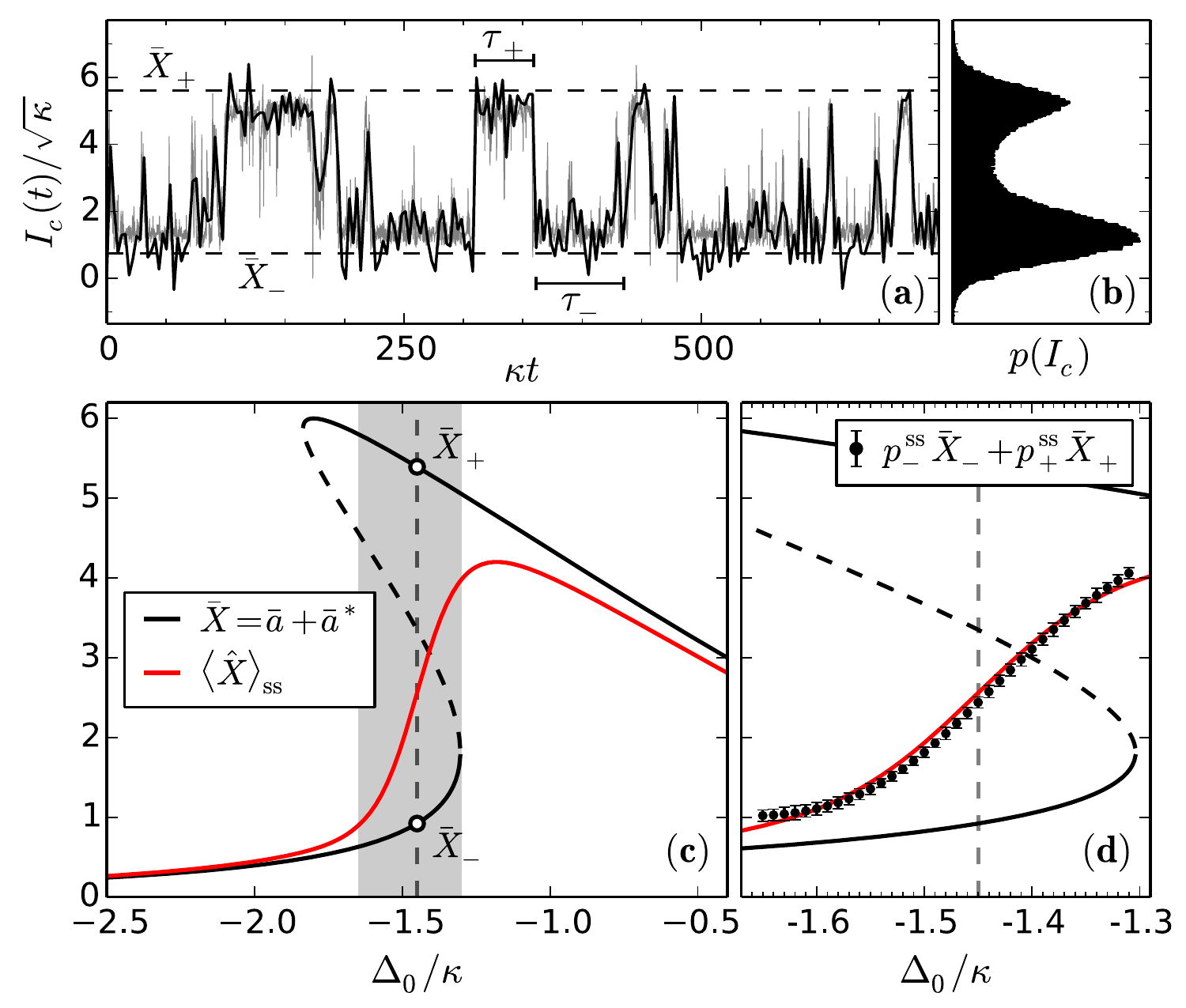}
\caption{(Color online) \textit{Noise-activated switching in a bistable
optomechanical system (OMS)}. (a) Homodyne photocurrent $I_c(t)$ for a
representative quantum trajectory. The OMS switches between two bistable states
that are close to the two stable solutions $\bar{X}_\pm$ (dashed lines)
of the nonlinear mean-field equations (MFEs) \eqref{eq:MFE}. From the time trace
$I_c(t)$ the residence times $\tau_\pm$ can be extracted. We show both the
conditioned expectation value of the amplitude quadrature $\langle \hat{X}(t)
\rangle_c$ (grey solid) and the homodyne photocurrent $I_c(t)$ after applying a
low-pass filter (black solid). (b) From a sufficiently long trajectory we can
obtain the probability distribution $p(I_c)$ of the filtered homodyne
photocurrent whose double-peak structure is a signature of the bistable
behavior. (c) Stable $\bar{X}_\pm$ (black solid) and unstable (black dashed)
solutions to the MFEs (\ref{eq:MFE}) as a function of the bare detuning
$\Delta_0$. We indicate the stable states (circles) between which the system
shown in (a) and (b) switches. The figure also shows the steady-state
expectation value $\langle \hat X \rangle_\text{ss}$ (red solid) interpolating
between the bistable solutions $\bar{X}_\pm$. (d) A blow-up of the region marked
grey in panel (c). Additionally, we plot the weighted average of the mean-field
solutions $p_-^\text{ss}\bar{X}_- + p_+^\text{ss}\bar{X}_+$ (black dots) where
the probabilities $p_\pm^\text{ss}$ are given by Eq.~\eqref{eq:ssprobability}.
The parameters are $\omega_m/\kappa = 5$, $\gamma_m/\kappa = 1/2$, $g_0/\kappa =
1/\sqrt{2}$, $\epsilon/\kappa = 1.5$, and $\Delta_0/\kappa = -1.45$ (a,b).}
\label{fig1}
\end{figure}

We investigate the dynamics of an OMS in a regime where the mechanical
resonator acts like an effective Kerr nonlinearity for the optical
mode \cite{Aldana2013}.  As a consequence the system can exhibit
optical bistability, a phenomenon characterized by the presence of two
stable mean-field states. In a semiclassical approximation the
steady-state amplitudes of the optical $\bar{a}$ and mechanical modes
$\bar{b}$ are obtained by solving the coupled mean-field equations
(MFEs)
\begin{equation}
\begin{aligned}
0 &= \left(i\Delta_0 - \frac{\kappa}{2} \right) \bar{a} +ig_0\bar{a} \left( \bar{b}+\bar{b}^{*}\right) + \epsilon,\\
0 &= -\left(i\omega_m + \frac{\gamma_m}{2}\right) \bar{b} + ig_0 |\bar{a}|^2\:.
\end{aligned}
\label{eq:MFE}
\end{equation}
An analysis of the nonlinear MFEs \eqref{eq:MFE} shows that the OMS undergoes a
bifurcation when the driving amplitude exceeds the threshold value
$\epsilon_\text{bif} = 3^{1/4} (\kappa^3\omega_m/18)^{1/2}/g_0$. As a
consequence three solutions for $\bar{a}$ exist in a certain range of negative
detuning $\Delta_0$. The two solutions $\bar{a}_\pm$ with the smallest and
largest amplitude $|\bar{a}|$ are stable and referred to as the upper and lower
branches of the bistable system.

Shot-noise fluctuations in the cavity drive will cause transitions between the
stable branches. This effect dubbed noise-activated switching has been
investigated e.g.~in the case of a Kerr medium theoretically
\cite{Rigo1997,Dykman2005,Dykman2007,Dykman2012,Peano2014} and experimentally
\cite{Kerckhoff2011}.

In Fig.~\ref{fig1}(a) we show the homodyne photocurrent $I_c(t)$ for a
representative quantum trajectory. We observe that the OMS switches between two
bistable states characterized by two different values of $I_c(t)$ and
corresponding approximately to $\sqrt{\kappa}\bar{X}_\pm$ where $\bar{X}_\pm =
\bar{a}_\pm + \bar{a}_\pm^{*} $. After applying a low-pass filter to the raw
quantum trajectory data we can extract the residence times $\tau_\pm$ from the
time trace $I_c(t)$. From a sufficiently long trajectory we obtain the
probability distribution $p(I_c)$ for the homodyne photocurrent, shown in
Fig.~\ref{fig1}(b). It features a double peak, a signature of
the bistable behavior.

In Fig.~\ref{fig1}(c) we show the mean-field amplitude quadrature, $\bar{X} =
\bar{a}+\bar{a}^{*}$, as function of the detuning $\Delta_0$ obtained from the
solutions to the nonlinear MFEs (\ref{eq:MFE}). We also calculate the
steady-state expectation value $\langle \hat X \rangle_\text{ss}$ from the QME
(\ref{eq:QME}) which interpolates between the two bistable solutions $\bar{X}_\pm$.

\begin{figure}
\includegraphics[width=\linewidth]{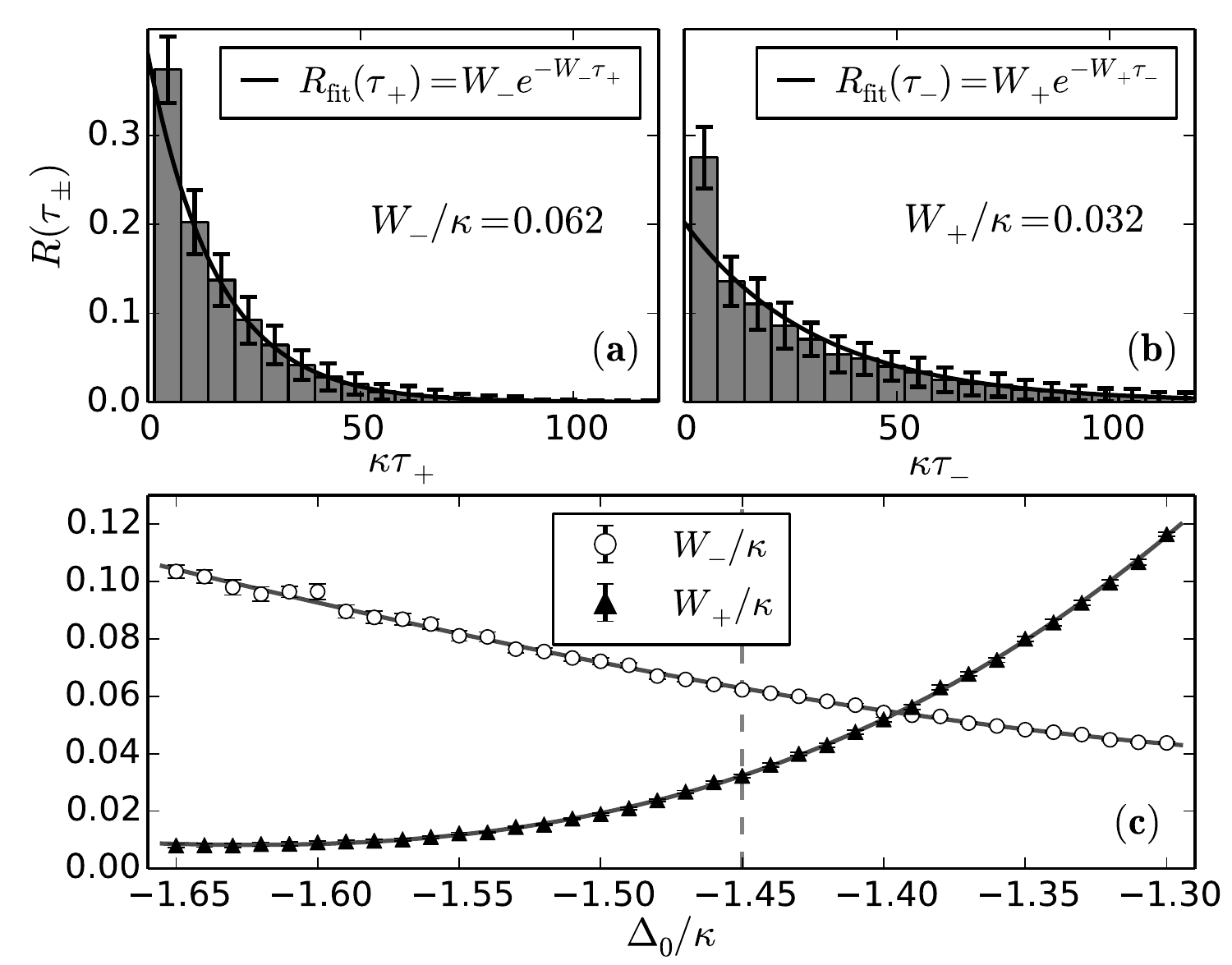}
\caption{(Color online) \textit{Residence time distributions and switching
rates}. (a) Histogram $R(\tau_+)$ of residence times in the upper branch
extracted from the quantum trajectory in Fig.~\ref{fig1}(a) with
statistical error bars.  The solid line is an exponential fit
$R_\text{fit}(\tau_+) = W_- e^{-W_-\tau_+}$ excluding the first bin.
(b) Same as (a) but for the residence times in the lower branch. We determine
the switching rate $W_+$ by fitting the histogram $R(\tau_-)$ with the
distribution $R_\text{fit}(\tau_-) = W_+ e^{-W_+\tau_-}$ . (c) Switching rates
$W_\pm$ as a function of $\Delta_0$. Parameters are identical to those in
Fig.~\ref{fig1} and with $\Delta_0/\kappa = -1.45$ (a,b).}
\label{fig2}
\end{figure}

Figures~\ref{fig2}(a) and \ref{fig2}(b) show histograms $R(\tau_\pm)$ of residence times
in the upper and lower branches, respectively, which we extracted from the
quantum trajectory shown in Fig.~\ref{fig1}(a) including statistical error
bars.  We fit the data with exponential distribution functions
$R_\text{fit}(\tau_\pm) = W_\mp \exp(-W_\mp\tau_\pm)$ and determine the
switching rates $W_\mp$ from the upper to the lower branch and vice versa
\footnote{The first bin of the residence time distributions $R(\tau_\pm)$
deviates from the exponential distribution $W_\pm e^{-W_\pm\tau_\mp}$. This is a
due to our definition of a switching event as the photocurrent $I_c$ crossing a
certain threshold value $I_\text{th}$. Fluctuations in each branch, noticeably
larger in the upper branch, can cause fake consecutive switching events. This effect
can be mitigated by applying a low-pass filter as shown in
Fig.~\ref{fig1}(a).}. In Fig.~\ref{fig2}(c) we plot the switching rates $W_\pm$
as a function of cavity detuning $\Delta_0$.

In steady state the probability to find the OMS in the upper or lower
branch, $p_\pm^\text{ss}$, is related to the switching rates via
\begin{equation}
p_\pm^\text{ss} = \frac{W_\pm}{W_+ + W_-}\:.
\label{eq:ssprobability}
\end{equation}
The probability $p_\pm^\text{ss}$ is the fraction of time spent by the system
in the upper and lower branch, respectively. It can be written as $T_\pm/(T_+ +
T_-)$,  where $T_\pm$ is the average residence time and is given by $T_\pm =
\int \tau_\pm R(\tau_\pm) d\tau_\pm = W _\mp^{-1}$.

If the fluctuations in each branch $\bar{a}_\pm$ are small compared to their
phase-space separation $|\bar{a}_+-\bar{a}_-|$, the average homodyne
photocurrent $I_\text{ss} = \text{E}[I_c(t)]$, or equivalently the steady-state
expectation value $\langle \hat{X} \rangle_\text{ss} =
I_\text{ss}/\sqrt{\kappa}$, is well approximated by the weighted average of the
mean-field solutions
\begin{equation}
I_\text{ss} \simeq \sqrt{\kappa} \left(p_-^\text{ss}\bar{X}_- + p_+^\text{ss}\bar{X}_+ \right)\:.
\label{eq:sshomodynephotocurrent}
\end{equation}
In Fig.~\ref{fig1}(d) we show a blow up of Fig.~\ref{fig1}(c) for detunings in
the bistable regime. Additionally, we also plot $p_-^\text{ss}\bar{X}_- +
p_+^\text{ss}\bar{X}_+$ where the probabilities $p_\pm^\text{ss}$ are given by
Eq.~\eqref{eq:ssprobability}. We see that the switching dynamics of bistable
OMS in this regime can be accurately captured by a two-state model.

\section{Two-state model with slowly and periodically modulated switching rates}
\label{sec:twostatemodel}

The influence of the periodic force (\ref{eq:force}) on the switching dynamics
can be described with a two-state rate equation model
\begin{align}
\dot{p}_\pm(t) & = \pm W_+(t) p_-(t) \mp W_-(t) p_+(t)\nonumber \\
& = -W(t) p_\pm(t) + W_{\pm}(t)
\label{eq:rateequation}
\end{align}
where $p_\pm(t)$ is the probability for the system to be in the vicinity of the
branch $\bar{a}_\pm$ satisfying $p_+ + p_- = 1$, $W_\pm(t)$ are the
time-dependent switching rates, and $W(t) = W_+(t) + W_-(t)$.

For a mechanical forcing that is slow on the time scale of intra-branch
fluctuations, i.e.~$\Omega \ll \kappa,\omega_m$, the influence of $\hat{H}_F$
can be reduced to an adiabatic change of the resonator equilibrium position that
is given by $ 2(g_1/\omega_m) \sin(\Omega t)$ in units of its zero-point
amplitude. This leads to a slow variation of the cavity detuning $\Delta_0 +
2(g_0 g_1/\omega_m) \sin(\Omega t)$ and will only affect the long-time dynamics of
the optical mode, i.e.~the switching behavior, by modulating the switching rates
\begin{equation}
W_\pm(t) = W_\pm^0 + W_\pm^1 \sin(\Omega t)\:.
\label{eq:transitionrates}
\end{equation}
Here, $W_\pm^0$ denote the switching rates in absence of the external force
$g_1=0$ and, assuming that for a weak force the switching rates depend linearly
on the detuning, we have
\begin{equation}
W_\pm^1 =  \frac{2g_0g_1}{\omega_m} \frac{\partial W_\pm^0}{\partial \Delta_0}\:.
\label{eq:Wplusminus1}
\end{equation}

The steady-state solution to the rate equation \eqref{eq:rateequation} for
periodic switching rates $W_\pm(t)$ with period $T_\Omega = 2\pi/\Omega$ is
itself periodic and given by \cite{Loefstedt1994}
\begin{equation}
\begin{aligned}
p_\pm (t) = & \frac{1}{1-e^{-\overline{W} T_\Omega}} \int_0^{T_\Omega} dt' W_\pm(t-t') \\
& \;\times e^{-\overline{W} t'} \exp\left[-\int_{t-t'}^{t} \delta W(t'') dt''\right]
\label{eq:genericsolutionRE}
\end{aligned}
\end{equation}
with $\overline{W} = \int_0^{T_\Omega} W(t) dt/T_\Omega$ and $\delta W(t) = W(t)
- \overline{W}$. For the transitions rates $W_\pm(t)$ in
Eq.~\eqref{eq:transitionrates}, $\overline{W} = W_+^0 + W_-^0$ and $\delta W(t)
= (W_+^1 + W_-^1)\sin(\Omega t)$. Expanding the exponential in
Eq.~\eqref{eq:genericsolutionRE} and neglecting higher harmonics, we obtain in
the limit $|W_+^1 + W_-^1| \ll \Omega$ the long-time solution
\begin{equation}
p_\pm (t) \simeq \frac{W_\pm^0}{\overline{W}} \pm \frac{W_+^1 W_-^0 - W_-^1
W_+^0}{\overline{W} \sqrt{\overline{W}^2 + {\Omega}^2}} \sin\left( \Omega t -
\phi\right)
\label{eq:timedependantprob}
\end{equation}
where $\phi = \arctan\left(\Omega/\overline{W} \right)$. The first term in
Eq.~\eqref{eq:timedependantprob} corresponds to $p_\pm^\text{ss}$, the
steady-state probability to find the system in the upper or lower branch in
absence of the external force. The second term is a slow periodic modulation of
these probabilities and we will use them to characterize the influence of an
external force on the homodyne photocurrent $I_c(t)$.

\section{Detection of weak periodic forces with a bistable optomechanical system}
\label{sec:forcedetection}

We will now analyze our force detection scheme
by examining the output spectral density of the homodyne photocurrent
$S_{II}^\text{out}(\omega)$. In brief, the spectral density is the sum of two
contributions, a noise background and a signal contribution,
\begin{equation}
S_{II}^\text{out}(\omega) = S_{II}^\text{noise}(\omega) + S_{II}^\text{signal}(\omega)\:.
\label{eq:spectraldensity}
\end{equation}
The noise background $S_{II}^\text{noise}(\omega)$ quantifies the
power per unit bandwidth of the noise interfering with detection at
frequency $\omega$. As we will show, in our detection scheme, the main contribution to
$S_{II}^\text{noise}(\omega)$ at low frequencies originates from the incoherent
switching of $I_c(t)$ between the two stable branches. A weak harmonic force
with frequency $\Omega$ produces a coherent modulation of the homodyne
photocurrent with amplitude $I(\Omega)$ and thus contributes a delta peak to the
spectral density
\begin{equation}
S_{II}^\text{signal}(\omega) = \frac{\pi}{2} I(\Omega)^2 \left[ \delta(\omega-\Omega) + \delta(\omega+\Omega) \right]\:.
\label{eq:signalpeak}
\end{equation}
For a finite sampling time $T$ one expects the signal peak height to be
$S_{II}^\text{signal}(\Omega) = \pi I(\Omega)^2/(2\Delta\omega)$ where
$\Delta\omega=2\pi/T$ is the finite frequency resolution of the spectral
density.

We will use two quantities to quantify the amplification and the sensitivity of our
proposed detector scheme. The first one is the ratio $I(\Omega)/g_1$ which relates
the modulation amplitude of the homodyne photocurrent $I(\Omega)$ (output
signal amplitude) to the forcing amplitude $g_1$ (input signal amplitude). This
ratio characterizes amplification with a dimensionless power gain
\begin{equation}
\mathcal{G}(\Omega) = \kappa \left(\frac{I(\Omega)}{g_1} \right)^2
\end{equation}
expressing the ratio of the signal output power $\propto I(\Omega)^2$ to the
signal input power $\propto g_1^2$. To quantify the sensitivity of our scheme
we will use the signal-to-noise ratio (SNR) defined as
\begin{equation}
\text{SNR} = \frac{\displaystyle
\frac{1}{\Delta\omega}\int_{\Omega-\Delta\omega/2}^{\Omega+\Delta\omega/2}
S_{II}^\text{out}(\omega) d\omega}{S_{II}^\text{noise}(\Omega)}\:.
\end{equation}
For a sufficiently long sampling time $T$, the noise background
$S_{II}^\text{noise}(\omega)$ is approximately constant over the frequency
window $\Delta\omega = 2\pi/T$. Thus, $\text{SNR} =
S_{II}^\text{signal}(\Omega)/S_{II}^\text{noise}(\Omega) + 1$, i.e.~the SNR
depends only on the ratio of the output signal and the noise background power at
the signal frequency $\Omega$.

Our two-state rate equation model allows us to find approximate expressions for
the noise spectral density $S_{II}^\text{noise}$ and signal amplitude
$I(\Omega)$. We will compare these analytical results to quantum trajectory
simulations below. Using the gain $\mathcal{G}(\Omega)$ and SNR to characterize
our detection scheme we will be able to compare its performance to force
detection with an OMS in the linear regime. We will derive analytical
expressions for the modulation amplitude $I_\text{lin}(\Omega)$, the power gain
$\mathcal{G}_\text{lin}$, and the noise background
$S_{II,\text{lin}}^\text{noise}$. We then express
$S_{II,\text{lin}}^\text{noise}$ as a function of the power gain
$\mathcal{G}_\text{lin}$ and the OMS parameters $\omega_m$, $\kappa$, and
$\gamma_m$ so we can compare the sensitivity of the two different schemes,
bistable OMS and linear OMS, at fixed power gain.

\subsection{Two-state approximation for the output spectral density}

\begin{figure}
\includegraphics[width=\linewidth]{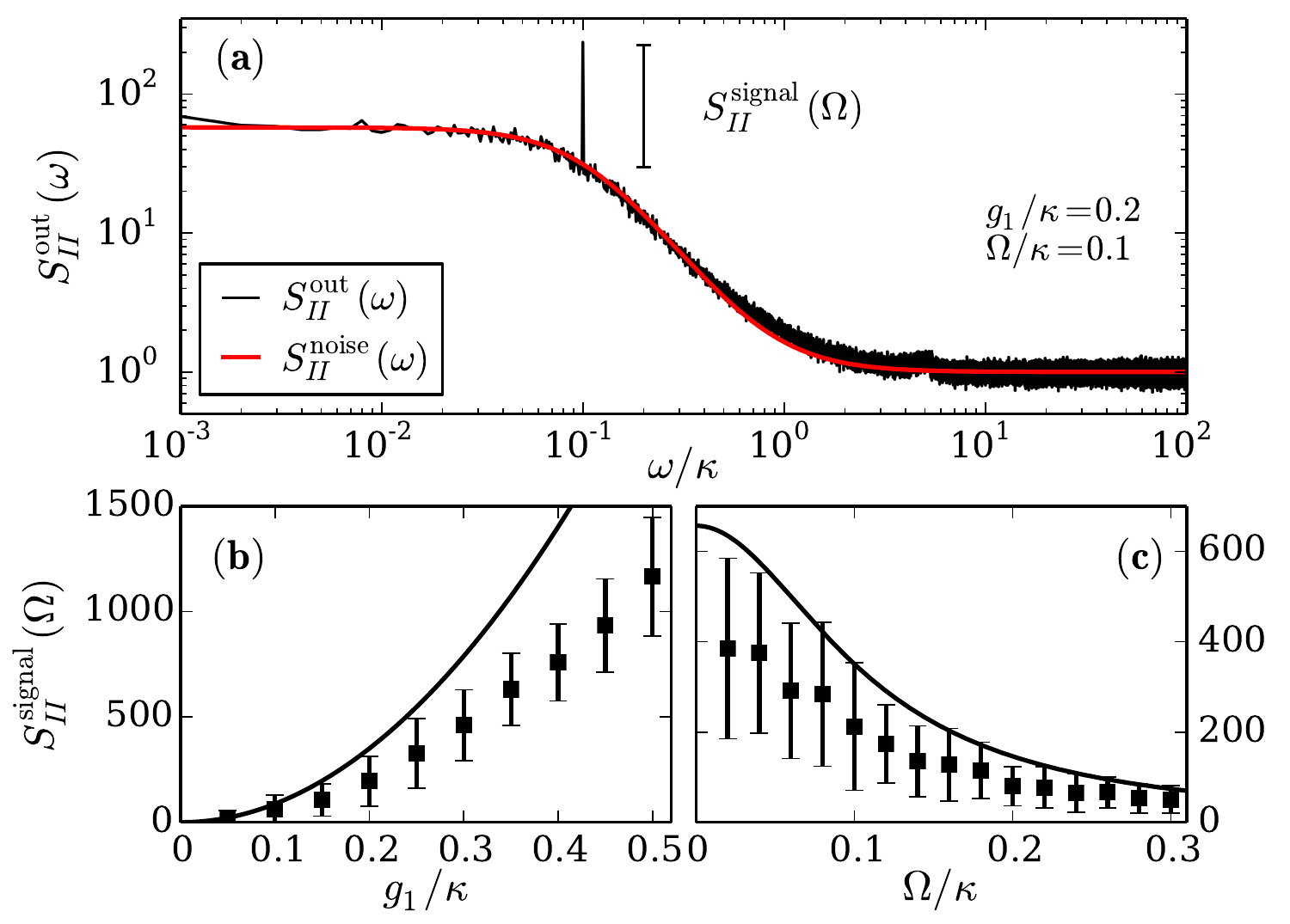}
\caption{(Color online) \textit{Detection of weak force with a bistable OMS}.
(a) Spectral density for the homodyne photocurrent $S^\text{out}_{II}(\omega)$
in presence of a weak external force on the mechanical oscillator (black).
The spectral density features a noise background and a signal peak. At small
frequencies the noise background $S^\text{noise}_{II}(\omega)$ (red) can be
approximated by a Lorentzian of width $\overline{W}$ at zero frequency,
Eq.~\eqref{eq:noisebackground}. (b) and (c) Signal peak height
$S^\text{signal}_{II}(\Omega) =
S_{II}^\text{out}(\Omega)-S_{II}^\text{noise}(\Omega)$ as a function of forcing
amplitude $g_1$ (b) and forcing frequency $\Omega$ (c). Black squares are
quantum trajectory simulations with statistical error bars. Black lines are
analytical results based on the two-state rate equation model, Eq.~\eqref{eq:peakheight}, as discussed in the main text. The parameters are the same as in
Fig.~\ref{fig1} but for $\Delta_0/\kappa = -1.4$. The weak external mechanical
force has a frequency $\Omega/\kappa = 0.1$ (a,b) and an amplitude $g_1/\kappa
= 0.2$ (a,c).
The spectral density for each pair of parameters $(\Omega,g_1)$ is obtained from
an average over hundred spectra with a frequency resolution $\Delta\omega =
10^{-3}\kappa$.}
\label{fig3}
\end{figure}

Describing the switching dynamics within the two-state rate equation model
allows us to find analytic expressions for the low-frequency part of the output
spectral density $S_{II}^\text{out}(\omega)$. As stated above,
Eq.~\eqref{eq:spectraldensity}, $S_{II}^\text{out}(\omega)$ can be separated
into a noise background $S_{II}^\text{noise}(\omega)$ and the signal part
$S_{II}^\text{signal}(\omega)$.

In absence of the external force incoherent switching causes
autocorrelations of the homodyne photocurrent to decay exponentially on a time
scale $\overline{W}^{-1}$. We find the autocorrelation function
(up to an irrelevant constant $I_\text{ss}^2$) is given by
\begin{equation}
\text{E} \left[ I_c(t+\tau) I_c(t) \right] = e^{-\overline{W}|\tau|} \kappa
p_+^\text{ss} p_-^\text{ss} (\bar{X}_+ - \bar{X}_-)^2 + \delta(\tau)\:.
\end{equation}
The second term stems from the shot noise of the local oscillator. The first
term is proportional to the steady-state variance
$\text{Var}(\hat{X})_\text{ss}=\langle\hat{X}^2\rangle_\text{ss}-\langle\hat{X}\rangle_\text{ss}^2
\simeq p_+^\text{ss} p_-^\text{ss}(\bar{X}_+-\bar{X}_-)^2$. Calculating
$\text{Var}(\hat{X})_\text{ss}$ from the QME~\eqref{eq:QME}, we find that this
two-state approximation overestimates the variance in the presence of
appreciable intra-branch fluctuations around mean-field solutions.
In fact, the noise background is smaller and more accurately
given by
\begin{equation}
S_{II}^\text{noise}(\omega) =2\kappa
\text{Var}(\textbf{:}\hat{X}\textbf{:})_\text{ss}
\frac{\overline{W}}{\overline{W}^2 + \omega^2} + 1\:,
\label{eq:noisebackground}
\end{equation}
where
$\text{Var}(\textbf{:}\hat{X}\textbf{:})_\text{ss}=\text{Var}(\hat{X})_\text{ss}-1$
is the normally-ordered variance of the amplitude quadrature, the colon
denoting normal ordering of the optical creation and annihilation operators.
Equation~\eqref{eq:noisebackground} satisfies the constraint that the total
power of the homodyne photocurrent minus the shot-noise contribution must
satisfy \cite{Wiseman1993},
$\int[S_{II}^\text{noise}(\omega)-1]\frac{d\omega}{2\pi}=\kappa\text{Var}(\textbf{:}\hat{X}\textbf{:})_\text{ss}$.
The noise spectrum consists of a shot noise contribution and a Lorentzian
centered at zero frequency with a half width at half maximum given by
$\overline{W}$.

Equation \eqref{eq:timedependantprob} allows us to find an approximate expression
for the signal part $S_{II}^\text{signal}$ to the output
spectral density. In the long-time limit a periodic time-dependence of the
probability $p_\pm(t)$ yields a periodically modulated average homodyne
photocurrent
\begin{equation}
\begin{aligned}
\text{E}[I_c(t)] & =\sqrt{\kappa}\left[p_+(t)\bar{X}_++p_-(t)\bar{X}_-\right]
\\ & = I_\text{ss}+I(\Omega) \sin(\Omega t-\phi)
\end{aligned}
\end{equation}
with the modulation amplitude in two-state approximation
\begin{equation}
I(\Omega) = \sqrt{\kappa}\left( \bar{X}_+ - \bar{X}_- \right) \frac{W_+^1 W_-^0
- W_-^1 W_+^0}{\overline{W} \sqrt{{\overline{W}}^2 + \Omega^2}}\:.
\end{equation}
The relationship between the average steady-state homodyne photocurrent
$I_\text{ss}=\sqrt{\kappa}\langle\hat{X}\rangle_\text{ss}$, the probabilities
$p_\pm^\text{ss}$, and the transitions rates $W_\pm^i$ given by
Eqs.~\eqref{eq:ssprobability}, \eqref{eq:sshomodynephotocurrent}, and
\eqref{eq:Wplusminus1} provide a direct interpretation of $I(\Omega)$. The
zero-frequency expression $I(0)=(2g_1g_0/\omega_m) (\partial
I_\text{ss}/\partial\Delta_0)$ is the linear response of $I_\text{ss}$ to a
change in the detuning $\Delta_0$. The prefactor $2g_1/\omega_m$ is the
zero-frequency response of the mechanical oscillator, i.e.~the change in the
mechanical equilibrium position (in units of its zero-point amplitude) caused
by a static force with amplitude $g_1$. This displacement leads to a
change of the cavity detuning $\Delta_0$ by $g_0(2g_1/\omega_m)$.
Relaxation of a bistable OMS at rate $\overline{W}$ causes
an attenuation of this response at finite frequencies $\Omega$,
\begin{equation}
I(\Omega) = \frac{2g_0g_1}{\omega_m} \sqrt{\kappa}\frac{\partial
\langle\hat{X}\rangle_\text{ss}}{\partial\Delta_0}
\frac{\overline{W}}{\sqrt{\overline{W}^2+\Omega^2}}\:.
\label{eq:modulationamplitude}
\end{equation}
As stated in Eq.~\eqref{eq:signalpeak}, the signal contributes a delta peak to
the spectral density since the autocorrelation function of the homodyne
photocurrent is dominated by periodic modulation in the limit $\tau \gg
\overline{W}^{-1}$, and hence factorizes, $\text{E}[I_c(t+\tau)I_c(t)] =
\text{E}[I_c(t+\tau)]\text{E}[I_c(t)]$.
For a finite frequency resolution $\Delta\omega$, 
\begin{equation}
S_{II}^\text{signal}(\Omega) = \frac{\pi \kappa}{2\Delta\omega}
\left(\frac{2g_1g_0}{\omega_m}\frac{\partial\langle\hat{X}\rangle_\text{ss}}{\partial\Delta_0}\right)^2
\frac{\overline{W}^2}{\overline{W}^2 + \Omega^2}.
\label{eq:peakheight}
\end{equation}

In Fig.~\ref{fig3}(a) we plot the spectral density for the homodyne
photocurrent $S^\text{out}_{II}(\omega)$ in the presence of a weak external force.
An average over hundred spectra is shown. The spectral density features a
low-frequency Lorentzian noise background whose frequency dependence agrees
very well with our two-state approximation $S_{II}^\text{noise}(\omega)$,
Eq.~\eqref{eq:noisebackground}. The height of the signal peak relative to the
noise level, $S_{II}^\text{signal}(\Omega) = S_{II}^\text{out}(\Omega) -
S_{II}^\text{noise}(\Omega)$, is obtained for a range of forcing amplitudes
$g_1$ and forcing frequencies $\Omega$. Comparing these quantum trajectory
simulations to Eq.~\eqref{eq:peakheight}, we find that
$S_{II}^\text{signal}(\Omega)$ exhibits the correct quadratic dependence on the
forcing amplitude $g_1$ and Lorentzian dependence on the forcing frequency
$\Omega$. The modulation amplitude $I(\Omega)$ is about 20\% smaller than
expected. We suspect that this quantitive disagreement is due to the large
amplitude of intra-branch fluctuations reaching a considerable fraction of the
inter-branch separation and the fact that the linear approximation to the
modulation of switching rates \eqref{eq:Wplusminus1} is only satisfied for the
smaller values of $g_1$ in Fig.~\ref{fig3}.

The expected power gain of a bistable OMS is
\begin{equation}
\mathcal{G}(\Omega) = \left(\frac{2g_0\kappa}{\omega_m}
\frac{\partial\langle\hat{X}\rangle_\text{ss}}{\partial\Delta_0}\right)^2
\frac{\overline{W}^2}{\overline{W}^2 + \Omega^2}\:.
\label{eq:powergain}
\end{equation}
We notice that amplification occurs over a bandwidth given by the switching
rate $\overline{W}$. As can be seen in Fig.~\ref{fig1}(c), the slope
$\partial\langle\hat{X}\rangle_\text{ss}/\partial\Delta_0$ in the center of the
bistable region is approximately proportional to the difference between the two
mean-field solutions $\bar{X}_+-\bar{X}_-$. As a consequence, a large
difference in the cavity output fields leads to a strongly amplified homodyne
signal. If the cavity is driven further away from bifurcation, the slope
increases, but the switching rate $\overline{W}$ decreases. Thus, the gain can
be made larger at the expense of reducing the bandwidth. For low signal
frequency, $\Omega\lesssim\overline{W}$, for which the shot-noise contribution
to the noise background $S_{II}^\text{noise}$ is negligible, the SNR is
independent of $\Omega$,
\begin{equation}
\text{SNR} \simeq \pi \frac{\overline{W}}{\Delta\omega}
\left(\frac{g_1g_0}{\omega_m}\right)^2
\frac{\left(\partial\langle\hat{X}\rangle_\text{ss}/\partial\Delta_0\right)^2}{\text{Var}(\textbf{:}\hat{X}\textbf{:})_\text{ss}}
+ 1\:,
\label{eq:SNR}
\end{equation}
with $\text{Var}(\textbf{:}\hat{X}\textbf{:})_\text{ss}$ and
$(\partial\langle\hat{X}\rangle_\text{ss}/\partial\Delta_0)^2$ obtained from
Eq.~\eqref{eq:QME}.
These two quantities have a similar dependence on the detuning $\Delta_0$ and
reach their maximum at an optimal value of $\Delta_0$ in the center of the
bistable region. As a consequence, both the SNR and the gain $\mathcal{G}$ are maximal.

\begin{figure}
\includegraphics[width=\linewidth]{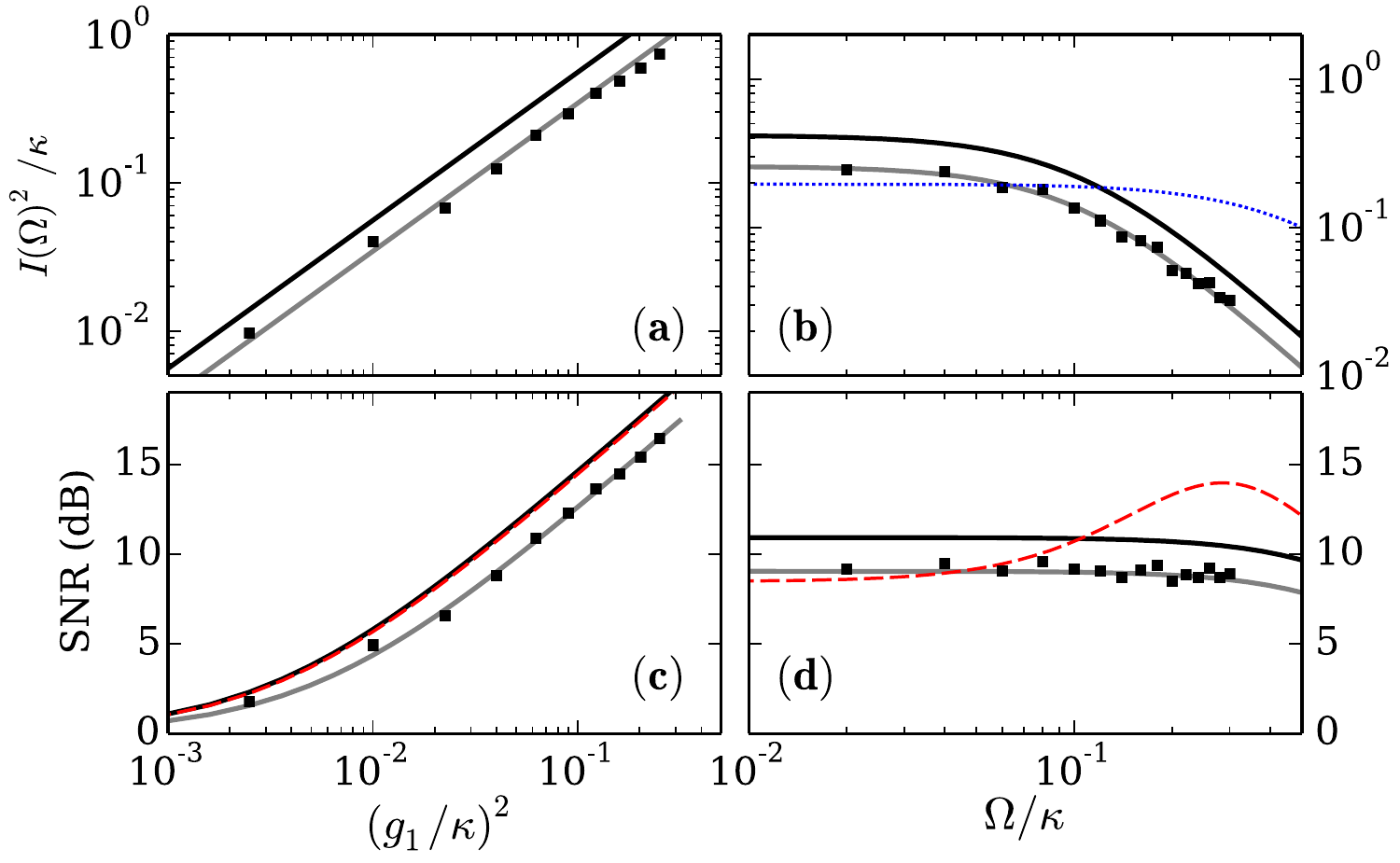}
\caption{(Color online) \textit{Power gain and signal-to-noise ratio (SNR)}.
Signal output power (a,b) and SNR (c,d) as function of the signal input power
(a,c) and signal frequency (b,d). The expected values of $I(\Omega)^2$ and the
SNR according to the two-state model discussed in the main text (black line)
are compared to quantum trajectory results shown in Fig.~\ref{fig3} (black
squares). Grey lines are a fit to the data indicating that the power gain
$\mathcal{G}(\Omega)$ and the SNR have the correct dependence on the signal
input power and signal frequency. The observed power gain has a value about
40\% smaller than expected. In panel (b), the dotted blue line indicates the
result for the largest possible power gain of an OMS operating in the linear
regime $\mathcal{G}_\text{lin}^\text{(max)}$, Eq.~\eqref{eq:maxpowergain}. In
panels (c) and (d), the dashed red line indicates the SNR for an OMS in the
linear regime operating at the same power gain (extracted from the quantum
trajectory results) and obtained from Eq.~\eqref{eq:SNRlinear}. The parameters
are identical to Fig.~\ref{fig3}, with an external forcing frequency
$\Omega/\kappa = 0.1$ (a,c) and amplitude $g_1/\kappa = 0.2$ (b,d).}
\label{fig4}
\end{figure}

Figure~\ref{fig4} shows the dimensionless signal output power
$I(\Omega)^2/\kappa$ (a,b) and SNR (c,d) as a function of the signal input power
$(g_1/\kappa)^2$ (a,c) and signal frequency $\Omega$ (b,d). We compare results
from quantum trajectory simulations and from our two-state rate equation model.
In panel (a) we see that the bistable OMS exihibits nearly constant power gain
for small forcing amplitudes $g_1$. In panel (b) we observe that its detection
bandwidth is in good agreement with predictions of the two-state model and given
by the switching rate $\overline{W}$. As expected, the SNR is approximately
constant over the detection bandwidth as can be seen in panel (c).

\subsection{Force detection with an OMS in the linear regime}

In the linear regime the dissipative dynamics of an OMS, including the noise
and signal spectral densities of its output field quadratures, can be obtained
exactly from the input-output formalism \cite{Gardiner1985,Clerk2010}. The
linear regime is characterized by a small optomechanical coupling rate,
$g_0\ll\kappa,\omega_m$, and a cavity driven to a coherent state with large
amplitude $|\bar{a}|\gg 1$. Under these conditions, the radiation-pressure
interaction can be approximated by a bilinear interaction, with an enhanced
coupling rate $g=g_0|\bar{a}|$, between the resonator position,
$\hat{b}+\hat{b}^\dagger$, and the amplitude quadrature,
$\hat{a}+\hat{a}^\dagger$. The static shift of the resonator position results
in an effective cavity detuning $\Delta = \Delta_0+g_0(\bar{b}+\bar{b}^{*})$. A
displacement of the mechanical resonator imprints a phase shift on the output
light field, which is best probed by driving the cavity on resonance,
$\Delta=0$, and by measuring the \textit{phase} quadrature at the output
\cite{Aspelmeyer2013}.

Analogous to Eq.~\eqref{eq:modulationamplitude} we find an expression for
the amplitude modulation $I_\text{lin}$ and the spectral density
$S_{II,\text{lin}}^\text{signal}$ of the phase quadrature in homodyne
detection due to the force
\begin{equation}
\begin{gathered}
I_\text{lin}(\Omega) = \sqrt{\mathcal{G}_\text{lin}(\Omega)}
\frac{g_1}{\sqrt{\kappa}},\\
S_{II,\text{lin}}^\text{signal}(\omega)= \frac{\pi}{2} I_\text{lin}^2(\Omega)
\left[ \delta(\omega-\Omega)+\delta(\omega+\Omega) \right]\:.
\end{gathered}
\label{eq:signallinear}
\end{equation}
Here, the equivalent power gain at frequency $\omega$ for an OMS in the linear
regime reads
\begin{equation}
\mathcal{G}_{\text{lin}}(\omega) = \left| 2 g \kappa \chi_c(\omega) \left[ \chi_m(\omega)-\chi_m^{*}(\omega)\right] \right|^2\:,
\end{equation}
with $\chi_c(\omega) = (\kappa/2-i\omega)^{-1}$ the cavity susceptibility and
$\chi_m(\omega) = [\gamma_m/2+i(\omega_m-\omega)]^{-1}$ the mechanical
susceptibility. The zero-frequency response can be written as $I_\text{lin}(0)
= (2g_0g_1/\omega_m) [\partial_\Delta(\sqrt{\kappa}\bar{I})]_{\Delta=0}$,
i.e.~the product of a shift of the cavity detuning caused by a static force
with amplitude $g_1$ and the derivative with respect to $\Delta$ of the average
homodyne photocurrent, $\sqrt{\kappa}\bar{I}$, where
$\bar{I}=-i(\bar{a}-\bar{a}^{*})$ is the mean-field value of the optical phase
quadrature and $\bar{a}=\epsilon/(\kappa/2-i\Delta)$.
At low frequency, $\omega\ll\kappa,\omega_m$, the power gain is approximately constant,
\begin{equation}
\mathcal{G}_\text{lin}(\omega) = \left(\frac{2g_0\kappa}{\omega_m}\left[\frac{\partial\bar{I}}{\partial\Delta}\right]_{\Delta=0}\right)^2\:,
\label{eq:powergainlinear}
\end{equation}
which is analogous to Eq.~\eqref{eq:powergain}.

The low-frequency power gain, Eq.~\eqref{eq:powergainlinear}, can as well be
expressed as $\mathcal{G}_\text{lin}(\omega) = (8g_0/\omega_m)^2 \bar{n}$, and is
proportional to the average cavity occupation on resonance,
$\bar{n}=|\bar{a}|^2=4(\epsilon/\kappa)^2$. An OMS can only operate in the
linear regime below bifurcation, $\epsilon<\epsilon_\text{bif}$, that is for a
cavity occupation below the critical value $n_\text{bif} =
2\kappa\omega_m/(3\sqrt{3}g_0^2)$. As a consequence, the power gain cannot be
made arbitrarily large and the maximal gain has the universal value
\begin{equation}
  \mathcal{G}_\text{lin}^\text{(max)}(\omega) 
\simeq \frac{128}{3\sqrt{3}} \frac{\kappa}{\omega_m}\:.
\label{eq:maxpowergain}
\end{equation}

The spectral density of the noise interfering with the detection of a force signal far
from the mechanical resonance, $|\omega-\omega_m|\gg\gamma_m$, referred
back to the input signal is \cite{Clerk2010,Aspelmeyer2013}
\begin{equation}
\begin{aligned}
\frac{S_{II,\text{lin}}^\text{noise}(\omega)}{\mathcal{G}_\text{lin}(\omega)} = &
\frac{1}{\mathcal{G}_\text{lin}(\omega)}
+\mathcal{G}_\text{lin}(\omega) \frac{\left(\omega_m^2-\omega^2\right)^2}{16 \kappa^2\omega_m^2} \\
& + \left( n_\text{th}+\frac{1}{2}\right) \frac{\gamma_m}{\kappa} \frac{\omega^2+\omega_m^2}{2\omega_m^2}\:.
\end{aligned}
\label{eq:noisebackgroundlinear}
\end{equation}
Equation (\ref{eq:noisebackgroundlinear}) expresses the total measurement noise
as fluctuations in the forcing amplitude and has three contributions.
The first term is the imprecision noise due to the shot noise of the local
oscillator. The second term is the back-action noise or radiation-pressure shot
noise. The last term originates from thermal and quantum fluctuations of the
resonator position.

At each frequency $\omega$, there is an optimal gain
$\mathcal{G}_\text{lin}^\text{(opt)}(\omega)=2\kappa|\chi_m(\omega)-\chi_m^{*}(-\omega)|$
for which the measurement noise is minimal and the SNR maximal. In the limit of
small frequencies, the optimal gain is then
\begin{equation}
\mathcal{G}_\text{lin}^{\text{(opt)}} \simeq \frac{4\kappa}{\omega_m}\:.
\end{equation}
The low-frequency noise level for the optimal gain and a mechanical
resonator coupled to a zero-temperature bath ($n_\text{th}=0$),
$S_{II,\text{lin}}^\text{noise} \simeq 2+ \gamma_m/\omega_m$, is minimal. This is
commonly referred to as the standard quantum limit (SQL) of force (or position)
detection. At the SQL the back-action noise and the imprecision noise are both
equal to the shot-noise term.

\subsection{Comparison of bistable and linear detection}

An OMS in the regime of optical bistability exhibits a power gain
$\mathcal{G}$ much larger than the gain $\mathcal{G}_\text{lin}$ of a
linear OMS. The low-frequency expressions for the power gain of a
bistable or linear OMS, Eqs.~\eqref{eq:powergain} and
\eqref{eq:powergainlinear}, depend on the coefficients
$(\partial\langle\hat{X}\rangle_\text{ss}/\partial\Delta_0)^2$ and
$(\partial\bar{I}/\partial\Delta)^2$, respectively. These coefficients
characterize the response of the steady-state value of the optical
amplitude and phase quadratures, respectively, to a change in the
detuning. The second coefficient is proportional to the average cavity
occupation, which is limited by $\bar{n}<n_\text{bif}$.  For a
bistable OMS,
$\partial\langle\hat{X}\rangle_\text{ss}/\partial\Delta_0$ is
proportional to the difference between the mean-field solutions
$\bar{X}_+-\bar{X}_-$ and can exceed $\partial\bar{I}/\partial\Delta$
far from the bifurcation. For small signal frequency
$\Omega<\overline{W}$, the gain $\mathcal{G}(\Omega)$ is much larger
than the optimal gain $\mathcal{G}_\text{lin}^\text{(opt)}(\Omega)$ at
which the SQL applies, and can even be larger than
$\mathcal{G}_\text{lin}^\text{(max)}(\Omega)$, i.e.~the maximal gain
for a linear OMS below bifurcation.

Figure~\ref{fig4}(b) shows the dimensionless signal output power
$I(\Omega)^2/\kappa$ as a function of the signal frequency $\Omega$ obtained
from quantum trajectory simulations and from the two-state model. In addition,
we indicate the results corresponding to a linear OMS operating at its  maximal
power gain $\mathcal{G}_\text{lin}^\text{(max)}$. Note that
$\mathcal{G}(\Omega) > \mathcal{G}_\text{lin}^\text{(max)}(\Omega)$ within the
detection bandwidth, i.e.~for signal frequencies $\Omega \lesssim \overline{W}$.

As a consequence of the large gain
$\mathcal{G}\gg\mathcal{G}_\text{lin}^\text{(opt)}$,
the measurement noise $S_{II}^\text{noise}$ unavoidably exceeds the
SQL value that applies to an OMS in the linear regime,
$S_{II,\text{lin}}^\text{noise} \simeq 2 + \gamma_m/\omega_m$.
Thus, instead of comparing the sensitivity of our scheme to a linear OMS
operating at the SQL, we compare it to the sensitivity of a linear OMS with
identical gain. The SNR of a linear OMS can be expressed as a function of its power gain
$\mathcal{G}_\text{lin}$ to compare it to results of quantum
trajectory simulations. From Eqs.~\eqref{eq:signallinear} and
\eqref{eq:noisebackgroundlinear}, we obtain, for small signal frequencies
$\Omega\ll\kappa,\omega_m$ and $n_\text{th}=0$,
\begin{equation}
\text{SNR}= \frac{\pi g_1^2}{2\Delta\omega\kappa}
\left[\frac{1}{\mathcal{G}_\text{lin}(\Omega)} +
\mathcal{G}_\text{lin}(\Omega) \frac{\omega_m^2}{16\kappa^2} +
\frac{\gamma_m}{4\kappa}\right]^{-1}.
\label{eq:SNRlinear}
\end{equation}

In Fig.~\ref{fig4}, we plot the SNR of a bistable OMS as function of the signal
input power $(g_1/\kappa)^2$ (c) and signal frequency $\Omega$ (d). In
addition, we plot the SNR of a linear OMS with identical parameters $\omega_m$,
$\gamma_m$, and $\kappa$ and operating at the same gain
$\mathcal{G}_\text{lin}(\Omega) = \mathcal{G}(\Omega)$, where
$\mathcal{G}(\Omega)$ is extracted from quantum trajectory simulations. An
important feature can be observed in panels (b) and (d) at signal frequencies
in the detection bandwidth, $\Omega\lesssim\overline{W}$. The power gain of
the bistable OMS exceeds $\mathcal{G}_\text{lin}^\text{(max)}$, while
the SNR is still comparable to what is expected for a linear OMS with
equal gain. Our results therefore indicate that large-gain force detection
with an OMS can be realized beyond bifurcation, while preserving a sensitivity
that is comparable to an equivalent linear OMS.

\section{Conclusion}
\label{sec:conclusion}

We have proposed bistable optomechanical systems as detectors of weak
harmonic forces. An external mechanical force modulates the cavity
frequency and thus the switching rates between the stable branches. A
large difference in the respective optical output fields will thus
lead to a strong amplification of the weak signal. The noise-induced
switching dynamics in the presence of a harmonic force is described by
a two-state rate equation model with periodically modulated switching
rates. Using this model, we have calculated the output signal and
noise spectral density relevant to homodyne detection of the optical
field and compared them to quantum trajectory simulations. Finally, we
have also compared the power gain and signal-to-noise ratio of our
detection scheme to those of an optomechanical system in the linear
regime. We find that a potentially larger gain can be achieved for
low-frequency force signals while preserving comparable force
detection sensitivity.  
These results point out a new direction for
the use of optomechanical devices exhibiting an appreciable
single-photon coupling rate for sensing applications requiring strong
amplification.

\begin{acknowledgments}
We would like to acknowledge interesting discussions with
G. Str\"ubi. This work was financially supported by the Swiss SNF
and the NCCR Quantum Science and Technology.
\end{acknowledgments}

\end{document}